\documentclass[11pt,english,a4paper]{article}
\usepackage{graphicx}
\usepackage{amstext}
\usepackage{caption}
\usepackage{etoolbox}
\usepackage{makeidx}
\include{epsf}
\usepackage{amsfonts}
\usepackage{authblk} 
\usepackage{sectsty}
\usepackage{amsmath,amssymb,epsfig}
\usepackage{amscd}
\usepackage{amsthm}
\usepackage{mathrsfs}
\usepackage{dsfont}
\usepackage[applemac]{inputenc}
\usepackage[english]{babel}
\usepackage{enumitem} 
\usepackage[]{latexsym}
\usepackage{caption}
\usepackage{subfigure}
\usepackage{hyperref}
\usepackage{blindtext}
\usepackage{verbatim}
\usepackage{cancel}
\usepackage{color}
\usepackage{mathtools}
\newcommand*{\affaddr}[1]{#1}
\newcommand*{\affmark}[1][*]{\textsuperscript{#1}}

\newtheorem*{proof*}{Proof}

\newcommand{\be}{\begin{equation}}

\newcommand{\ee}{\end{equation}}
\def\beqa{\begin{eqnarray}}
\def\eeqa{\end{eqnarray}}
\def\bean{\begin{eqnarray*}}
\def\eean{\end{eqnarray*}}

\renewenvironment{thebibliography}[1]
         {\section*{References}\frenchspacing\small
          \begin{list}{[\arabic{enumi}]}
         {\usecounter{enumi}\parsep=2pt\topsep 0pt
         \settowidth{\labelwidth}{[#1]}
         \leftmargin=\labelwidth\advance\leftmargin\labelsep
         \rightmargin=0pt\itemsep=1pt\sloppy}}{\end{list}}

 \numberwithin{equation}{section}

\input xy
\xyoption{all}

\usepackage[normalem]{ulem}

\newcommand{\ket}[1]{\left| #1 \right\rangle}

\newcommand{\vev}[1]{\left\langle #1 \right\rangle}

\newcommand{\braket}[2]{\left\langle \vphantom {#1 #2} #1 \hphantom{|} \right| \left. \vphantom {#1 #2} #2 \right\rangle}

\newcommand{\braopket}[3]{\left\langle \vphantom {#1 #2 #3} #1 \hphantom{|} \right| #2 \left| \hphantom{|} \vphantom {#1 #2 #3} #3 \right\rangle}

\newcommand{\ba}{\begin{eqnarray}}
\newcommand{\ea}{\end{eqnarray}}

\newcommand{\ov}{\hat{v}}
\newcommand{\oH}{\hat{H}_g}

\newcommand{\oC}{\hat{C}}
\newcommand{\oE}[1]{\widehat{e^{#1 i \lambda b}}}
\newcommand{\oep}{\oE{+}}
\newcommand{\oem}{\oE{-}}
\newcommand{\oepm}{\oE{\pm}}

\newcommand{\osin}{\widehat{\sin{(\lambda b)}}}
\newcommand{\ocos}{\widehat{\cos{(\lambda b)}}}
\newcommand{\of}[1]{f\BR{\ov #1}}

\newcommand{\oj}{\hat{j}_z}
\newcommand{\okx}{\hat{k}_x}
\newcommand{\oky}{\hat{k}_y}
\newcommand{\okp}{\hat{k}_+}
\newcommand{\okm}{\hat{k}_-}
\newcommand{\okpm}{\hat{k}_\pm}
\newcommand{\okmp}{\hat{k}_\mp}
\newcommand{\casi}{\hat{\mathscr{C}}}

\newcommand{\comm}[2]{\left[#1,#2\right]}

\newcommand{\BS}[1]{\left[ #1 \right]}
\newcommand{\BR}[1]{\left( #1 \right)}

\newcommand{\comme}[1]{\comm{#1}{\oepm}}

\newcommand{\osqva}[1]{\sqrt{\BR{\ov #1}^2 - \tilde{v}_m^2}}
\newcommand{\osqv}{\sqrt{\ov^2 - \tilde{v}_m^2}}

\newcommand{\osqma}[1]{\sqrt{\BR{2 \lambda m #1}^2 -\tilde{v}_m^2}}

\newcommand{\sqp}[1]{\hat R_{+ #1}}
\newcommand{\sqm}[1]{\hat R_{- #1}}

\newcommand{\sq}{\hat R_0}
\newcommand{\os}{\hat{s}}
\newcommand{\oc}{\hat{c}}
\newcommand{\lmket}{\ket{2 \lambda m}}

\newcommand{\Cf}{\ocos \, \ov \, \osin + \osin \, \ov  \, \ocos}
\newcommand{\Hf}{\osin \, \ov \, \osin}

\title{\textbf{\textsf{Renormalisation with SU$(1,1)$ coherent states on the LQC Hilbert space}}\vspace{0.35cm}}

\author{
\textsf{Norbert Bodendorfer\affmark[1]\footnote{\texttt{norbert.bodendorfer@physik.uni-r.de}}, Dennis Wuhrer\affmark[1]\footnote{\texttt{dennis.wuhrer@uni-konstanz.de}}}\\
\affaddr{\affmark[1]\textsf{Institute for Theoretical Physics, University of Regensburg,}}\\
\affaddr{\textsf{93040 Regensburg, Germany}}\vspace{-0.5cm}
}

\begin{document}

\maketitle

\begin{abstract}
\textsf{We present an analytic computation of an explicit renormalisation group flow for cosmological states in loop quantum gravity. 
A key ingredient in our analysis are Perelomov coherent states for the Lie group SU$(1,1)$ whose representation spaces are embedded into the standard loop quantum cosmology (LQC) Hilbert space. 
The SU$(1,1)$ group structure enters our analysis by considering a classical set of phase space functions that generates the Lie algebra $\mathfrak{su}(1,1)$.
We implement this Poisson algebra as operators on the LQC Hilbert space in a non-anomalous way. 
This task requires a rather involved ordering choice, whose existence is one of the main results of the paper. 
As a consequence, we can transfer recently established results on coarse graining cosmological states from direct quantisations of the above Poisson algebra to the standard LQC Hilbert space and full theory embeddings thereof. 
We explicitly discuss how the $\mathfrak{su}(1,1)$ representation spaces used in this latter approach are embedded into the LQC Hilbert space and how the $\mathfrak{su}(1,1)$ representation label sets a lower cut-off for the loop quantum gravity spins (= U$(1)$ representation labels in LQC). Our results provide an explicit example of a non-trivial renormalisation group flow with a scale set by the $\mathfrak{su}(1,1)$ representation label and interpreted as the minimally resolved geometric scale.  }
\end{abstract}

\section{Introduction}

Loop quantum gravity \cite{ThiemannModernCanonicalQuantum, RovelliBook2} is a non-perturbative approach to a quantum theory of the gravitational field. At its core are quantisation techniques similar to those of lattice gauge theory, but augmented to apply to background-independent theories. The key step in this procedure is to perform a quantisation of the gravitational field in terms of connection variables such that the gravitational degrees of freedom are represented as (non-regular) lattices on which the quantum dynamics acts. The representation labels of the involved gauge groups, in the standard formulation SU$(2)$ spins $j_\text{SU$(2)$}$, turn out to specify (some of the) the geometric properties of such lattices, in particular their physical proper size. This leads to the notion of a quantum geometry, a priori referring to spatial slices in canonical quantisations and space-times in path-integral formulations. For simplicity, let us consider a linear scaling of ``size'' with $j_\text{SU$(2)$}$, as will be the case in this paper, e.g. via the volume operator in the loop quantum cosmology setting. 

For computation of dynamical processes and an eventual comparison to experiment, one is, as usual, faced with the problem of different possible states describing the same coarse (quantum) geometry. If one is far away from the Planck scale, such states may be represented equally well by many small spins on fine lattices, or few large ones on coarse lattices, as far as coarse observables are concerned. These descriptions should be connected via a renormalisation group flow that renormalises the operators involved in the description. The study of such flows has received increasing attention in recent years in the loop quantum gravity literature, see e.g. \cite{MarkopoulouCoarseGrainingIn, OecklRENORMALIZATIONFORSPIN, LivineCouplingOfSpacetime, DittrichCoarseGrainingMethods, BahrHolonomySpinFoam, BahrOnBackgroundIndependent, BahrHypercuboidalRenormalizationIn, BahrNumericalEvidenceFor, CarrozzaFlowingInGroup, BahrRenormalizationOfSymmetry, DittrichCoarseGrainingFlow, BodendorferStateRefinementsAnd, BodendorferCoarseGrainingAs, ThiemannRenormalisationReview}. As a result of the complexity of the involved analytical problem, explicit results are however scarce. Rather, for dynamical computations, one usually works in the limit of large $j_\text{SU$(2)$}$ for which convenient asymptotic formula exist and defers the renormalisation problem to a later stage. While this typically leads to the correct semiclassical limit for curvatures much lower (and discretisation much larger) than the Planck scale, it is unclear what kind of a low spin formulation, if any, this would correspond to. As low spins are expected to be relevant in the high curvature regime where quantum gravity effects should be important, see e.g. \cite{HanSpinfoamsNearA} for a recent treatment, this question is rather pressing.   

In this paper, we are going to tackle this problem in a simplified setting that considers quantum states representing spatially flat, homogeneous and isotropic cosmology. Recent proposals \cite{GielenCosmologyFromGroup, OritiEmergentFriedmannDynamics, BodendorferStateRefinementsAnd} approximate such states as product states on $N$ identical (fiducial) cells, each corresponding to a copy of a Hilbert space that represents a single quantum cosmology restricted to one cell. The main technical simplification for coarse graining is that in this approximation, interactions between different cells can be neglected, leading to an effectively 0+1 dimensional problem when considering the possible interactions that may arise along the renormalisation group flow. We specifically note that in \cite{BodendorferStateRefinementsAnd}, the Hilbert space encoding the cosmological degrees of freedom has precisely the structure of the current paper, i.e. many copies of the LQC Hilbert space, so that our results apply to this full theory embedding of LQC.

A key ingredient in our analysis is a recent study \cite{BodendorferCoarseGrainingAs} of coarse graining in quantum cosmological models based on an $\mathfrak{su}(1,1)$ Lie algebra structure \cite{BojowaldDynamicalCoherentStates, BojraDynamicsForA, LivineGroupTheoreticalQuantization, BenAchourThiemannComplexifierIn} that allows to exactly compute a coarse graining flow under the assumption that the involved operators, the so-called CVH algebra (due to Complexifier-Volume-Hamiltonian, see below), are isomorphic to the generators of $\mathfrak{su}(1,1)$. While \cite{BodendorferCoarseGrainingAs} started from a classical Poisson algebra with this property, we are going to construct such operators directly on the loop quantum cosmology (LQC) Hilbert space in this paper. The $\mathfrak{su}(1,1)$ structure will then immediately yield an explicit and non-trivial renormalisation group flow under a change of scale, i.e. the transition from many small to few large spins. In this process, the $\mathfrak{su}(1,1)$ representation label $j$ functions as a lower cutoff for the involved U(1)-analogues of the SU(2) spins $j_\text{SU$(2)$}$ in loop quantum cosmology. 

The available SU$(1,1)$ Lie group structure allows us to use Perelomov coherent states \cite{PerelomovCoherentStatesFor}, which are a key ingredient in our discussion. Perelomov coherent states are a certain class of coherent states that live in the group representation spaces and can be constructed for a large class of groups. As such, they are useful wherever physical quantities reflect a group structure. Prominent examples include the quantisation of angular momentum operators \cite{PerelomovBook} and wavelets \cite{AntoineTwoDimensionalWavelets}.  

While we expect the methods developed in this paper to be applicable more generally, we focus here on the simplest possible cosmological setting, i.e. homogeneous, isotropic, and spatially flat. Throughout this paper, we only consider the gravitational part of the Hamiltonian constraint. While one may regard this as a toy model for a more general investigation including matter fields, such a system has a physical realisation when non-rotating dust is used to deparametrise the time direction \cite{BrownDustAsStandard, SwiezewskiOnTheProperties}. In this case, the gravitational part of the Hamiltonian constraint becomes the true Hamiltonian of the gravitational degrees of freedom. The error estimate for neglecting renormalisation that is performed section \ref{sec:RvNR} is thus strictly valid only for this case, but the general mechanism is expected to apply also to other treatments of the matter sector.

This paper is organised as follows:\\
\noindent Basic concepts of loop quantum cosmology are recalled in section \ref{sec:LQC}. Section \ref{sec:RecallCVH} reviews \cite{BodendorferCoarseGrainingAs} and the associated main idea of using the $\mathfrak{su}(1,1)$ structure of the CVH algebra for coarse graining. Our main result, an explicit realisation of the CVH algebra on the LQC Hilbert space and the implied renormalisation group flow, is presented in section \ref{sec:Main}. We conclude in section \ref{sec:Conclusion} and briefly survey the representation theory of $\mathfrak{su}(1,1)$ in the appendix.

\section{Basics of the LQC Hilbert space} \label{sec:LQC}

In this section, we briefly review the Hilbert space structure of loop quantum cosmology to the extend necessary for this paper. Seminal papers on the subject include \cite{BojowaldAbsenceOfSingularity, AshtekarMathematicalStructureOf, AshtekarQuantumNatureOf}, see \cite{AshtekarLoopQuantumCosmology, SinghLoopQuantumCosmologyABrief} for reviews. We consider spatially flat, homogeneous and isotropic cosmology where the gravitational sector is described by the canonical pair $\{b, v\}=1$, where $v$ is the signed spatial volume and $b$ is proportional to the mean curvature. We work in units where $\hbar = 12 \pi G = c=1$.

The aim of LQC is to quantise a cosmological model while mimicking key steps from full loop quantum gravity. The result should be considered as an ``inspired model'' unless one refers to a precise embedding of such a model into a full theory context, see e.g. \cite{AlesciANewPerspective, BIII, BVI, OritiEmergentFriedmannDynamics}. To avoid unnecessary technicalities and references to full loop quantum gravity, we introduce LQC as the synthesis of a spatially flat, homogeneous and isotropic quantum cosmological model in the presence of a spatial volume quantised in integer multiples of a fundamental scale $\lambda >0$. It follows that wave functions in the volume representation have support only on $\lambda \mathbb Z$ and the natural scalar product reads
\be
	\braket{\Psi_1}{\Psi_2} = \sum_{v \in \mathbb \lambda Z} \overline{\Psi_1(v)}\Psi_2(v) \text{.}
\ee
Hence, there cannot be an operator corresponding to $b$ as it would act as a derivative on a discontinuous function. Rather, the shift operators
\be
	\widehat {e^{i \lambda n b}}, ~n \in \mathbb Z 
\ee
have a well defined action on basis states $\ket{v}$ as 
\be
	\widehat {e^{\pm i \lambda n b}} \ket{v} = \ket{v \pm n \lambda} .
\ee
and are self-adjoint. 

One is therefore forced to regularise operators corresponding to $b$ or its powers via such exponentials, an often adopted\footnote{We will also make this choice, as it allows us to proceed analytically. We note that different choices are known to produce different physics, see e.g. \cite{AssanioussiEmergentDeSitter}, but will not investigate how this affects the current discussion.} choice being $b \mapsto \widehat{\sin(\lambda b)/\lambda}$. 
Such a replacement is referred to as a polymerisation and is analogous to using holonomies around closed loops instead of field strengths in lattice gauge theory.  We note that there is, at least so far and in this simple model, no continuum limit implied that would remove the correction terms $\mathcal O(\lambda^2 b^3)$. Rather, such terms should be interpreted as higher derivative quantum corrections to the effective action that are suppressed by the scale $\lambda$. In the spatially flat homogeneous and isotropic setting, this means that corrections become relevant once the matter energy density becomes close to $1/\lambda^2$. In LQC, one argues that $\lambda \approx 1$ is a natural choice \cite{AshtekarQuantumNatureOf}, leading to corrections close to the Planck curvature. 
One generically finds that cosmological singularities are resolved by such corrections, although exotic counterexamples can be constructed \cite{HellingHigherCurvatureCounter}. 

The above Hilbert space that we will denote as $\mathcal H_\text{LQC}$ can be identified with the square integrable functions on U$(1)$, where every integer value of $v/\lambda$ corresponds to a representation. 
Expansion in the $v$-basis can thus be understood as a Peter-Weyl decomposition of a function on U$(1)$, where the range of $b$ is compactified to $[0, 2\pi/\lambda)$. 
An extension to the Bohr compactification of the real line is possible where the main difference is that $n$ may be any real number, see e.g. \cite{AshtekarMathematicalStructureOf}. We will not consider this possibility here as the simplest choices for the dynamics preserve an evenly spaced lattice of $v$-values.

\section{Quantising the CVH algebra and coarse graining} \label{sec:RecallCVH}

In this section, we will review recent results on using the CVH algebra in the context of LQC. Seminal papers about using an $\mathfrak{su}(1,1)$ structure in LQC include \cite{BojowaldDynamicalCoherentStates, BojraDynamicsForA}. The algebra was directly quantised in \cite{LivineGroupTheoreticalQuantization, BenAchourThiemannComplexifierIn} using a LQC inspired regularisation preserving the $\mathfrak{su}(1,1)$ structure, following the seminal ideas of \cite{IshamTopologicalAndGlobal}. The usefulness of the associated SU$(1,1)$ coherent states in the context of coarse graining was pointed out in \cite{BodendorferCoarseGrainingAs}.

\subsection{CVH algebra and regularisation}

For classical spatially flat homogeneous and isotropic cosmology, the CVH algebra is formed by the so called complexifier $C = v b$ (owing its name to \cite{ThiemannComplexifierCoherentStates}), the spatial volume $v$, and the gravitational part of the Hamiltonian constraint $H_g = - \frac{1}{2} v b^2$, where $b$ is proportional to the mean curvature. 
The main feature of this set of phase space functions is that it forms a Poisson algebra isomorphic to $\mathfrak{su}(1,1)$ using the bracket $\{b, v\}=1$. Via the identification 
\be
	C = k_y, ~~~~ v = \frac{1}{2} \left( j_z + k_x \right), ~~~~ H_g = k_x - j_z
\ee 
or equivalently 
\be
 	j_z = v - \frac{1}{2}  H_g, ~~~~ k_x = v + \frac{1}{2}  H_g, ~~~~ k_y = C, \label{eq:Classsu11Mixed}
\ee
one finds 
\be
	\{ k_x, k_y\} = -j_z, ~~~ \{ k_y, j_z\} = k_x, ~~~ \{ j_z, k_x\} = k_y.
\ee
With the definition $k_\pm := k_x \pm i k_y$, the algebra reads
\be
	\{ k_+, k_-\} =  2 i j_z, ~~~ \{ j_z, k_\pm\} = \mp i k_\pm. \label{eq:CommRelPM}
\ee
We will use this latter form throughout the paper. 
A generalisation to the complete Hamiltonian constraint including a massless scalar field is possible, see e.g. \cite{BenAchourThiemannComplexifierIn}, although we will not consider it in this paper due to the value that the Casimir operator takes (see section \ref{sec:QuantCoarse} for more details). 
It should also be noted that other choices for the $\mathfrak{su}(1,1)$ generators are possible. Further details on the representation theory of $\mathfrak{su}(1,1)$ can be found in the appendix.
It will be relevant for later to also spell out the Poisson brackets of the CVH algebra in terms of $C$, $v$, and $H_g$:
\be
	\{v, H_g\} = C, ~~~~ \{C, v\} = v, ~~~~ \{ C, H_g\}= - H_g \text{.} \label{eq:CVHAlg}
\ee

For the intended application to loop quantum cosmology, it is necessary to consider the CVH algebra also for polymerised quantities. 
In a slight abuse of notation, we will again call them $C$, $v$, and $H_g$, as the classical quantities above do not appear any more in this paper. 
Due to holonomy corrections, $C$ and $H_g$ are changed to \cite{BenAchourThiemannComplexifierIn}
\be
	C = \frac{1}{2\lambda} v \sin{(2 \lambda b)}, ~~~~ H_g = - \frac{1}{2 \lambda^2} v \sin{(\lambda b)}^2 \text{,} \label{eq:HolonomyCorrections}
\ee
so that the classical quantities are obtained in the limit $\lambda \rightarrow 0$. The $\mathfrak{su}(1,1)$ algebra can be reproduced via the identification 
\begin{align}
	j_z = \frac{v}{2 \lambda}, ~~~~ k_x = \frac{1}{2 \lambda} \left( 4 \lambda^2 H_g + v \right), ~~~~ k_y = C \text{.} \label{eq:SU11Polym}
\end{align}
Using \eqref{eq:SU11Polym}, one finds that the classical Casimir operator $\mathfrak C = j_z^2 - k_x^2-k_y^2$ vanishes. In order to have access also to different representations of $\mathfrak{su}(1,1)$ in the quantum theory, which is needed for coarse graining, we augment \eqref{eq:SU11Polym} to a one-parameter family that yields different values for the Casimir operator, corresponding to different representations.
Suitable candidate regularisations that generalise \eqref{eq:SU11Polym} to such a one-parameter family were obtained in \cite{LivineGroupTheoreticalQuantization, BenAchourThiemannComplexifierIn} as
\be
	j_z = \frac{v}{2\lambda}, ~~~~ k_\pm = \frac{\sqrt{v^2-v_m^2}}{2 \lambda} e^{\pm 2 i \lambda b} \text{,} \label{eq:ClassAlgEasy}
\ee 
from which \eqref{eq:CommRelPM} can be easily verified. Here, $v_m$ is a free constant that can be identified with a minimal volume and will be relevant later in assigning suitable $\mathfrak{su}(1,1)$ representations. 
In fact, the classical Casimir operator now evaluates to
\be
	\mathfrak C = j_z^2 - k_x^2-k_y^2 = \frac{v_m^2}{4 \lambda^2} \text{,} \label{eq:CasClass}
\ee
which should be matched, at least to leading order, with the value of the Casimir operator as determined by the representation choice. 
The main technical task of this paper is to find a suitable operator ordering for \eqref{eq:ClassAlgEasy} so that \eqref{eq:CommRelPM} also holds via commutators. 

An obvious question at this point is why one should insist of implementing precisely this algebra non-anomalously at the quantum level, and not some other. The reason for doing so in this paper is purely technical: it gives us access to techniques using Perelomov coherent states for the Lie group SU$(1,1)$ and, as a consequence, allows us to compute analytically a non-trivial renormalisation group flow by applying the results of \cite{BodendorferCoarseGrainingAs}. In particular, we are not aware of a deeper physical reason for selecting this algebra.

\subsection{Group quantisation} \label{sec:QuantCoarse}

Rather than quantising on the LQC Hilbert space, it is straight forward to quantise our system by promoting $j, k_\pm$ to the generators of $\mathfrak{su}(1,1)$ on the standard group representation spaces, see the appendix for an overview of the relevant representation theory. The representation problem is thereby already solved, it only remains to pick a suitable subclass of $\mathfrak{su}(1,1)$ representations. In order to be able to transfer the ideas of \cite{BodendorferCoarseGrainingAs} and focus only on the gravitational sector, we choose representations from the discrete class with representation label $j \in \mathbb N/2$ and positive eigenvalues for $\oj$. For such a representation the Casimir operator takes the value $j(j-1)$. This suggests to identify 
\be
	j = \frac{v_m}{2 \lambda} \text{,} \label{eq:jvm}
\ee
reproducing \eqref{eq:CasClass} up to a subleading correction in $j$. As we will see in the following, the identification \eqref{eq:jvm} is precise for another large class of operators that we will be interested in for coarse graining, i.e. $2 \lambda j$ is the minimal eigenvalue of the volume operator. 

An interesting choice of quantum states is given by the normalised $\mathfrak{su}(1,1)$ Perelomov coherent states \cite{PerelomovCoherentStatesFor},
\be
	\ket{j, z} = (2L)^{j} \sum_{m=j}^\infty \sqrt{\binom{m+j-1}{m-j}} \, \frac{(z^1)^{m-j}}{(\bar{z}^0)^{m+j}} \, \ket{j,m} \label{eq:DefCoh}
\ee
which are characterised by the representation label $j \in \mathbb N/2$ and a spinor $z \in \mathbb C^2$, where we abbreviated $L = \frac12 (|z^0|^2-|z^1|^2)$. 
They have the property that the $\mathfrak{su}(1,1)$ action in any representation with label $j$ transfers directly to the spinor, which is in the defining representation:
\be
	U\ket{j,z} = \ket{j, U \cdot z} ~ \forall ~ U \in \text{SU}(1,1) \text{.} \label{eq:UonCS}
\ee
This property will later allow to relate the dynamics between finer and coarser scales labelled by $j$. The $\ket{j, z}$ also form an over-complete basis of the representation space with label $j$. However, for the purpose of coarse graining, we do not consider superpositions of coherent states with different labels.

\subsection{Coarse graining} \label{sec:CoarseGraining}

It was shown in \cite{BodendorferCoarseGrainingAs} that the coherent states \eqref{eq:DefCoh} allow for a natural coarse graining operation. One first notes that the expectation values of $\hat j_\alpha$, where $\hat j_\alpha$ is any element of $\{\hat j_z, \hat k_+, \hat k_-\}$, which all scale with the (proper) size of the system, factor into 
\be
	\braopket{j, z}{\hat j_\alpha}{j, z} =   j \cdot f_\alpha(z)
\ee
where $f_\alpha$ are three functions depending only on $z$. This suggests to interpret $j$ as an extensive scale of the system, while $z$ sets the intensive state, i.e. ratios of extensive quantities. We note that this is consistent with the classical interpretation of $j_z, k_\pm$ if $v_m$ also scales extensively, which is precisely the case for the identification \eqref{eq:jvm} along with the additional observation that $j$ is the minimal eigenvalue of $j_z = v/(2 \lambda)$. The coarse graining operation now looks as follows: We consider $N$ independent (= non-interacting) copies of our system, where the quantum state in each copy is given by \eqref{eq:DefCoh} with the same $j_0, z$ \footnote{Such a product state is well motivated in the cosmological application we have in mind and similar to a group field theory condensate approach to the topic \cite{GielenHomogeneousCosmologiesAs, OritiEmergentFriedmannDynamics}.}. The interpretation of $j$ as a scale\footnote{Let us carefully define our notion of scale for clarity. We consider the total universe to be of a fixed finite proper size, either obtained via a compact topology or using a fiducial cell. This system can be described either as a single cell (as before this footnote), or patched together out of many finer, but identical cells. When we talk about scale in the context of coarse graining (from now on), we always refer to one of those finer cells. By proper size, we mean proper volume, i.e. the volume density integrated over the cell. Extensive quantities are then properties of the fine cells whose magnitude grows with the proper volume of a fine cell as we increase the size of the fine cell, thereby inversely proportionally decreasing the number of the fine cells making up our universe.} in turn suggests that one may obtain the same physics\footnote{To be precise, by ``physics'' here we mean the expectation values of arbitrary powers of $\hat j_\alpha$ (which form a complete set of observables), see equation \eqref{eq:ConjectureFullPP}, as well as the probabilities to obtain eigenvalues of $\hat j_z$, see equation \eqref{eq:ProbPropagation}.} if one instead considers a single copy labeled by $j, z$, where $j = N j_0$. The following coarse graining map thus suggests itself:

\begin{center}
  \begin{tabular}{ l | c | c }
    
     & Fine description & Coarse description \\ \hline
    Quantum state & $\prod_{i=1}^N \ket{j_0,z}_i$ & $\ket{j, z}$ \\ \hline
    Operators & $ \hat j_{\alpha, 1}^{(j_0)} + \ldots + \hat j^{(j_0)}_{\alpha, N} $ & $\hat j^{(j)}_{\alpha}$ \\
    \hline
  \end{tabular}
\end{center}
We note that while $\ket{j_0,z}_i$ carries an index $i$ to remind us that it belongs to a certain cell in the fine description, all $N$ such states are identical, i.e. $\ket{j_0,z}_i = \ket{j_0,z}$, $i = 1, \ldots, N$. The bracketed superscript on the $\hat j^{(j_0)}_\alpha$ operators reminds us that these are the respective generators in the $\mathfrak{su}(1,1)$ representation $j_0$. We consider only the sum of the extensive fine operators as other linear combinations should not be captured at the coarse grained level.

The coarse graining map proposed above turns out to be {\it exactly} correct for the expectation values of any power of $\hat j_\alpha \in \{\hat j_z, \hat k_+, \hat k_-\}$ if one, as suggested above, compares the coarse grained operators to a sum of the corresponding operators at the non-coarse grained level as suggested by their extensive nature\footnote{The more general situation of mixed terms in $j_z, k_\pm$, linear combinations, as well as matrix elements, will be investigated in a future publication \cite{BHAddendum}.} (we drop the bracketed superscript on the $\hat j_\alpha$ here to avoid clutter):
\ba
	 \vev{\hat j_\alpha ^n}_j \label{eq:ConjectureFullPP} 
	 &=& \vev{\left( \hat j_{\alpha, 1} + \ldots + \hat j_{\alpha, N} \right)^n}_{j_0, 1, \ldots, N} \\
	&=& \sum_{\substack{ r_1, \ldots, r_{j} = 0  :\\ n = r_1 + \ldots + r_{j}}}^n \frac{n!}{r_1! r_2! \ldots r_{j}!} \vev{ \hat j_\alpha^{r_1}}_{j_0} \ldots \vev{\hat j_\alpha^{r_j}}_{j_0}  \nonumber \\    
	&=& \sum_{m=1}^{n} \frac{N!}{(N-m)!} \sum_{\substack{1 \leq k_1 \leq \ldots \leq k_{m}  :\\ n = k_1 + \ldots + k_{m}}} \frac{n!}{k_1! k_2! \ldots k_{m}!} \prod_{p=1}^{n} \frac{1}{(\#k_i=p)!} \vev{\hat j_\alpha^p}_{j_0}^{(\#k_i=p)} \text{.} \nonumber 
\ea
In the first line on the right hand side, the additional subscript on $\hat j_\alpha$ refers to one of the $N$ copies of the system, and the expectation value is taken in the product state of $N$ states with labels $j_0, z$. 
Equality of the left and right hand side of the first line is proven in the appendix of \cite{BodendorferCoarseGrainingAs}. 
The second line uses the multinomial theorem to split the expression into products of expectation values of powers of $\hat j_\alpha$. In the third line, another convenient form is given where terms with the same powers are collected. Here, $(\#k_i=p)$ is the number of $k_i$ which take the value $p$.
 
Furthermore, the eigenvalues and their probability distributions are exactly reproduced. In a single cell $i$ at the fine level (in representation $j_0$), the possible eigenvalues of $\hat j_z$ are $k_i = j_0 + \mathbb N_0$, and thus at the coarse grained level $k = N j_0 + \mathbb N_0$. These are also the eigenvalues that $\hat j_z$ can take in representation $j$. As one would have hoped, the probabilities $P_{j,k}$ to obtain eigenvalue $k$ in representation $j$ satisfy
\be
	P^{\text{coarse}}_{Nj_0,k} = \sum_{\substack{ k_1, \ldots, k_{N} = 0  :\\ k = k_1 + \ldots + k_{N}}} P_{j_0,k_1} \cdot P_{j_0,k_2} \cdot  \ldots  \cdot P_{j_0,k_N} \text{,} \label{eq:ProbPropagation}
\ee
where the right hand side is simply a sum over all possibilities to obtain a certain coarse configuration. 

If the dynamics is generated by a linear combination of $j_z, k_\pm$ (as will be the case for the gravitational part of the Hamiltonian constraint), it is also correctly reproduced due to \eqref{eq:UonCS}.  
In other words, the coarse graining operation commutes with computing time evolution.
We note that these results have been derived in \cite{BodendorferCoarseGrainingAs} using only the representation theory of $\mathfrak{su}(1,1)$ and the choice of states \eqref{eq:DefCoh}. Hence, they are independent of the application to quantum cosmology that we focus on in this paper and only require a classical Poisson algebra of extensive quantities that is isomorphic to $\mathfrak{su}(1,1)$.

\section{The CVH algebra on the LQC Hilbert space} \label{sec:Main}

In the previous section, we have recalled how group quantisation can be used to directly quantise a classical algebra that has been identified with a Lie algebra. The straight forward coarse graining properties of the Perelomov coherent states make such a quantisation particularly attractive and one would like to import it somehow to the standard LQC Hilbert space. There however, one does not start with quantising a classical algebra that has been identified with $\mathfrak{su}(1,1)$, but with operators $\hat v$, $\widehat {e^{i n \lambda b}}$ that correspond to terms which were used in the classical definition \eqref{eq:ClassAlgEasy} of the Lie algebra generators. One therefore needs to consider derived operators in a certain ordering that should be (partially) fixed by requiring that the operators reproduce the $\mathfrak{su}(1,1)$ commutation relations. In the following, we will present such an ordering choice and discuss how the corresponding $\mathfrak{su}(1,1)$ representation spaces can be identified as subspaces of the LQC Hilbert space. Related results for the light-like representation with label $j=0$ were recently reported in \cite{BenAchourProtectedSL2RSymmetry}. 

In the following, the basic commutation relations
\begin{align}
\label{eq:commve}
[\ov, \oepm] = \pm \lambda \oepm, ~ ~~~
[\ov, \osin] = - i \lambda \ocos, ~ ~~~
[\ov, \ocos] = i \lambda \osin
\end{align}
as well as
\begin{align}
	\comme{\of{}} =  \oepm \BS{\of{\pm \lambda}-\of{}} = \BS{\of{} -\of{\mp \lambda}} \oepm
\end{align}
will be repeatedly applied. Details of all computations, which are cumbersome but straight forward, can be found in \cite{WuhrerMasterarbeit}.

\subsection{Warmup: no minimal volume} \label{sec:NoMinVol}

Due to the rather cumbersome computation necessary to derive the general result, we first consider the simplifying choice that the classical minimal volume appearing in \eqref{eq:ClassAlgEasy} vanishes and highlight why it is necessary to go beyond this restriction. 

We start by choosing $j_z = \frac{\hat v}{2 \lambda}$ motivated by the wish of having at least one simple operator and the identification of a minimal volume eigenvalue by $v_\text{gap}:=2 \lambda j$. Other choices are possible as e.g. in \eqref{eq:Classsu11Mixed}. Our strategy is then to pick a simple regularisation of $\hat { H}_g$, derive $\hat C$ via the commutation relations following from \eqref{eq:CVHAlg}, and check the resulting operator for consistency with the other commutation relations.

Our trial ansatz for $\hat { H}_g$ is the symmetric ordering 
\begin{align} 
\oH = -\frac{1}{2 \lambda^2} \Hf .
\label{eq:HV}
\end{align}
\eqref{eq:CVHAlg} implies that $\hat C$ should satisfy 
\begin{align}
\oC = -i\comm{\ov}{\oH} &= \frac{1}{2 \lambda}\BS{\Cf}.
\end{align}
Using this definition, we compute
\begin{align}
\comm{\oC}{\ov} = i \BR{4 \lambda^2 \oH + \ov} \text{,} \label{eq:BracketCvNMV}
\end{align}
reproducing the second relation of \eqref{eq:SU11Polym}, as well as the second relation of \eqref{eq:CVHAlg} in the limit $\lambda \rightarrow 0$.
To verify that the algebra is really closed, we still need to calculate the commutator of $\oH$ with $\oC$. Using
\begin{align}
\comm{\osin}{\oC} = i \, \osin \ocos^2,
\end{align}
we find
\begin{align}
\comm{\oH}{\oC} = i \oH.
\end{align}
With these results, we can define the operators 
\begin{align}
\begin{split}
\oj = \frac{\ov}{2\lambda}, \qquad \okx = \frac{1}{2 \lambda}\BR{4 \lambda^2 \oH + \ov}, \qquad \oky = \oC, \qquad \okpm = \okx \pm i \oky, 
\end{split}
\label{eq:SU11opV}
\end{align}
as the quantum analogues of \eqref{eq:SU11Polym} which fulfil the $\mathfrak{su}(1,1)$ algebra 
\begin{align}
\begin{split}
\comm{\oj}{\okpm} = \pm \okpm, \qquad \comm{\okp}{\okm} = -2 \oj.
\end{split}
\end{align}

We should now study the properties of the representation we found, i.e. find the representation label $j$ and study how the representation space is embedded in the LQC Hilbert space. Let us first note that due to the action $\hat j_z \ket{j,m} = m \ket{j,m}$, the accessible volume eigenstates $\ket{v}$ (see section \ref{sec:LQC}) correspond to eigenvalues $2 \lambda m$, $m = j, j+1, j+2, \ldots$, i.e. $\ket{j, m}_{\mathfrak{su}(1,1)} = \ket{2 \lambda m}_\text{LQC}$. From now on, we will use the labels $\ket{\ldots}_\text{LQC}$ and $\ket{\ldots}_{\mathfrak{su}(1,1)}$ on the quantum states for clarity.

Next, we would like to fix $j$ by requiring $\hat k_- \ket{j, m=j}_{\mathfrak{su}(1,1)} = 0$ due to \eqref{eq:ActionK-}, i.e. the lowest eigenstate of $\hat j_z$ is annihilated by $\hat k_-$. For this, we compute the action of $\okx$ and $\oky$ on the volume eigenstate $\ket{2 \lambda m}_{\text{LQC}}$ in $\mathcal H_{\text{LQC}}$ as
\begin{align}
\begin{split}
\okx \lmket_{\text{LQC}} =\frac{1}{2}\BS{\BR{ m + \frac{1}{2}}\ket{2 \lambda (m+1) }_\text{LQC} + \BR{m - \frac{1}{2}} \ket{2 \lambda (m-1) }_\text{LQC}}
\end{split}
\label{eq:kxquant}
\end{align}
and
\begin{align}
\begin{split}
\oky \lmket_{\text{LQC}} =  \frac{1}{2 i}\BS{\BR{ m + \frac{1}{2}}\ket{2 \lambda (m+1) }_\text{LQC}  - \BR{ m - \frac{1}{2}} \ket{2 \lambda (m-1) }_\text{LQC}}
\end{split}
\label{eq:kyquant}
\end{align}
leading to
\begin{align}
\okpm \lmket_{\text{LQC}} = \BR{\okx \pm i \oky}\lmket_{\text{LQC}} = \frac{1}{2 \lambda}\BR{\ov \mp \lambda} \oE{\pm 2} \lmket_{\text{LQC}}.
\end{align}
$\okm \ket{2\lambda m}_{\text{LQC}} = (m-\frac{1}{2})  \ket{2\lambda (m-1)}_{\text{LQC}} = 0$ implies $m=\frac{1}{2}$ and due to $m=j$ for the lowest eigenstate $\ket{j, m=j}_{\mathfrak{su}(1,1)}$ of $\hat j_z$, we have $j=\frac{1}{2}$.
As a cross-check, we can explicitly compute the action of the Casimir operator as
\begin{align}
\casi \lmket_{\text{LQC}} = \BR{\oj^2 - \okx^2 - \oky^2}\lmket_{\text{LQC}} =  -\frac{1}{4} \lmket_{\text{LQC}}
\end{align}
which is consistent due to $j(j-1) = -1/4$ for $j=1/2$. 

We conclude that while we found a factor ordering for the CVH algebra that reproduces the correct commutation relations, we are restricted to the $j=1/2$ representation. For applications of coarse graining as discussed in section \ref{sec:CoarseGraining}, it is interesting to also find explicit realisations of the CVH algebra on $\mathcal H_{\text{LQC}}$ for all $j \in \mathbb N/2$. The observation of section \ref{sec:QuantCoarse}, cited from \cite{BenAchourThiemannComplexifierIn}, that there is a one-parameter family of classical Poisson algebras labelled by $v_m \sim j$ suggests that a similar one-parameter family may yield the representations for $j>1/2$. As we will show in the next subsection, this expectations turns out to be correct.

\subsection{Regularisation with minimal volume}

\renewcommand{\Cf}{\ocos  \osqv   \osin + \osin \osqv  \ocos}
\renewcommand{\Hf}{\osin  \osqv  \osin}

Inspired by \eqref{eq:ClassAlgEasy}, we again choose $\hat j_z = \frac{\hat v}{2 \lambda}$. For $\oH$, we make the ansatz 
\begin{align}
\oH = -\frac{1}{2 \lambda^2}\Hf , \label{eq:AnsatzHgMV}
\end{align}
where $\tilde{v}_m$ is a free constant (multiplied by an identity operator, which is dropped to avoid clutter). As we will see later, consistency of the algebra determines it to be $\tilde{v}_m = {v}_\text{gap} - \lambda$, where ${v}_\text{gap}=2 \lambda j$ is the minimal eigenvalue of $\hat v$ restricted to the $\mathfrak{su}(1,1)$ representation sub-space of the LQC Hilbert space (see below), and it thus includes a subtle quantum correction to the naive classical expectation \eqref{eq:ClassAlgEasy}. That is, the classical minimal volume $v_m$ appearing in \eqref{eq:ClassAlgEasy} should be identified with $\tilde v_m$ in the quantum theory, whereas the volume observable in the quantum theory is bounded from below by $ v_\text{gap} = \tilde v_m + \lambda$.

Proceeding as before, we calculate $\oC$ via the commutator of $\oH$ and $\ov$ as
\begin{align}
\oC = -i \comm{\ov}{\oH} = \frac{1}{2 \lambda}\BS{\ocos \osqv \osin + \osin \osqv \ocos} \text{.}
\end{align}
The next step consists in computing $\comm{\oC}{\ov}$ as
\begin{align}
\comm{\oC}{\ov} = i \BR{4 \lambda^2 \oH + \frac{1}{2}\BS{\osqva{+\lambda} + \osqva{-\lambda}}} \label{eq:CommCVMV}.
\end{align}
Comparison with \eqref{eq:BracketCvNMV} shows that the $\hat v$ term obtains corrections for non-zero $\tilde v_m$.  

As before, we are left with the commutator $\comm{\oH}{\oC}$. To simplify the calculation, we will introduce the shorthand notation
\begin{align}
\begin{split}
&\oc = \ocos, \qquad \os = \osin,  \qquad \hat R_n =  \osqva{+n\lambda} .
\end{split}
\label{eq:shorthand}
\end{align}
Straight forward but cumbersome computations yield
\begin{align}
\comm{\os\sq\os}{\oc\sq\os + \os\sq\oc} &= - \frac{1}{2 i} \left( {\sqp{1} - \sqm{1}}\os\sq\os + \os\sq\os\BR{\sqp{1} - \sqm{1}} \right) \nonumber \\
							     &= - \frac{1}{2i} \left( {\sqp{1}\os^2\sqp{1}-\sqm{1}\os^2\sqm{1}} \right) \text{,}
\end{align}
leading to 
\begin{align}
\comm{\oH}{\oC} &= \frac{1}{8i\lambda^3}\BR{\sqp{1}\os^2\sqp{1}-\sqm{1}\os^2\sqm{1}} \label{eq:CommHCMV2} \\
			  &= \frac{-i \ov}{4 \lambda^2}  + \frac{i}{32 \lambda^3}\BS{\oep\BR{\sqp{2}\sq - \sqm{2}\sq}\oep + \oem\BR{\sqp{2}\sq - \sqm{2}\sq}\oem} \text{.}  \nonumber
\end{align}
To calculate the commutator of $\oC$ and \eqref{eq:CommCVMV}, one first shows that
\begin{align}
\frac{1}{2}\comm{\sqp{1} + \sqm{1}}{\oC} = \frac{-i}{8 \lambda} \BS{\oep\BR{\sqp{2}\sq - \sqm{2}\sq}\oep + \oem\BR{\sqp{2}\sq - \sqm{2}\sq}\oem} \text{.}
\end{align}
Putting this result together with \eqref{eq:CommHCMV2} yields 
\begin{align}
\comm{\oC}{4 \lambda^2 \oH + \frac{1}{2}\BR{\sqp{1}+\sqm{1}}} = i \ov.
\end{align}
As a last step, we compute
\begin{align}
\comm{\ov}{4 \lambda^2 \oH + \frac{1}{2}\BR{\sqp{1}+\sqm{1}}} = \comm{\ov}{4 \lambda^2 \oH} = 4 \lambda^2 i \oC.
\end{align}
Analogously to \eqref{eq:SU11opV}, we are now in a position to define the $\mathfrak{su}(1,1)$ generators as 
\begin{align}
\okx = \frac{1}{2 \lambda}\BS{4 \lambda^2 \oH + \frac{1}{2}\BR{\sqp{1}+\sqm{1}}}, \qquad \oky = \oC, \qquad \oj = \frac{1}{2 \lambda} \ov, \qquad \okpm = \okx \pm i \oky \label{eq:RegVM}.
\end{align}
Our previous calculations imply that their algebra reproduces that of $\mathfrak{su}(1,1)$ as
\begin{align}
 \comm{\oj}{\okpm} = \pm \okpm \qquad \comm{\okp}{\okm} = -2 \oj \text{.}
\end{align}

Since $\oj$ did not change as compared to the previous subsection, the accessible volume eigenstates $\ket{v}_{\text{LQC}}$ still have the eigenvalues $2 \lambda m$, $m = j, j+1, j+2, \ldots$. To fix the representation, we again need to calculate the action of $\okpm$ on volume eigenstates. Using the intermediate results
\begin{align}
\okx \lmket_{\text{LQC}} &= \frac{1}{4\lambda} \BR{\osqma{+\lambda}\ket{2 \lambda (m+ 1)}_{\text{LQC}} + \osqma{-\lambda}\ket{2 \lambda (m-1)}_{\text{LQC}}} \label{eq:actionkx} \\
i \oky \lmket_{\text{LQC}} &= \frac{1}{4 \lambda}\BR{\osqma{+ \lambda}\ket{2 \lambda (m+1)}_{\text{LQC}} - \osqma{-\lambda}\ket{2 \lambda (m-1)}_{\text{LQC}}}, 
\end{align}
we obtain
\begin{align}
\begin{split}
\okpm \lmket_{\text{LQC}} &=  \frac{1}{2\lambda}\osqma{\pm \lambda}\ket{2 \lambda (m \pm 1)}_{\text{LQC}} \text{.} \label{eq:kpmActionMV}
\end{split}
\end{align}
The action of $\okm$ vanishes for $\tilde v_m = \pm \lambda (2m-1)$. As $m\stackrel{!}{=}j$ ($m$ should agree with $j$) in this case, we find using $ v_\text{gap} = \tilde v_m + \lambda$ that
\be
	 j =   \frac{1}{2 \lambda }\BR{\tilde{v}_m + \lambda }= \frac{ v_\text{gap}}{2 \lambda}
\ee
for the choice $ v_\text{gap} > \tilde v_m \geq 0$ and $j>0$. Again, we can confirm this result via the Casimir operator as
\begin{align}
\begin{split}
\casi \lmket_{\text{LQC}} =  \BR{  \frac{\tilde{v}_m^2}{4 \lambda^2} - \frac{1}{4}} \lmket_{\text{LQC}} = \frac{ v_\text{gap}}{2 \lambda} \left(\frac{ v_\text{gap}}{2 \lambda}-1 \right) \lmket_{\text{LQC}} \stackrel{!}{=}  j(j-1)\lmket_{\text{LQC}} 
\end{split}
\end{align}
For $j=1/2$ $\Leftrightarrow$ $ v_\text{gap} = \lambda$, we obtain the results from section \ref{sec:NoMinVol} as a cross-check.

\subsection{Embedding of the $\mathfrak{su}(1,1)$ representation spaces}

Let us now collect our results. As shown before, by regularising the CVH algebra as \eqref{eq:RegVM} on $\mathcal H_\text{LQC}$, we can correctly reproduce the $\mathfrak{su}(1,1)$ algebra. 
It follows from the representation theory of $\mathfrak{su}(1,1)$ that by acting with operators from the CVH algebra, in particular $\okp$,  on volume eigenstates with eigenvalue $\geq 2 \lambda j$, we obtain states with higher volume, but never go below the minimal volume $2 \lambda j$. 
Therefore, in a representation $j$, the support of wave functions in the $\mathfrak{su}(1,1)$ representation spaces embedded into $\mathcal H_\text{LQC}$ is restricted to $v \in 2 \lambda (j + \mathbb N_0)$. 
In particular, it follows that the dynamics generated by any linear combination of $\oj, \okpm$ preserves this subspace. This last observation is crucial for the viability of our coarse graining operation.

\subsection{Coarse graining} \label{subsec:CC}

The results obtained in this section, combined with \cite{BodendorferCoarseGrainingAs}, allow to define a coarse graining operation as follows. Using \eqref{eq:ConjectureFullPP}, we know that $N$ independent but identical copies of our quantum cosmological system described by a product state with labels $j_0, z$ in each cell can be equivalently described by a single copy with coherent state labelled by $j, z$ with $j=Nj_0$. The only requirement for this is that the coarse grained $\oj, \okpm$ satisfy the $\mathfrak{su}(1,1)$ algebra and are represented in the representation with label $j$. We established this before. 

It is of interest to compare the coarse graining flow derived in this section with that of section \ref{sec:CoarseGraining}. We first note that in section \ref{sec:RecallCVH}, we used a different quantisation procedure that quantises a classical Poisson algebra isomorphic to $\mathfrak{su}(1,1)$. Therefore, the $\mathfrak{su}(1,1)$ generators are immediately available on the Hilbert space and no assembling of them from other more fundamental operators was necessary as in section \ref{sec:Main}. This means that $\oj, \okpm$ are unambiguously defined, while their supposed constituents, such as $\ov$, $\osin$, and $v_m$ are not. The point of view taken in \cite{BodendorferCoarseGrainingAs} was that $v_m$ in \eqref{eq:ClassAlgEasy} (that is analogous to $\tilde v_m$ in section \ref{sec:Main}), also scales extensively with the system size $j$. This may be achieved by defining $v_m$ as the lowest eigenvalue of $\ov$, while, e.g., a definition over the Casimir operator eigenvalue $j(j-1)$ suggests a quantum correction to the extensive scaling when comparing with \eqref{eq:CasClass}. Such discrepancies are expected because there is no unique definition of operators that are not contained in the sub-algebra of observables that one (unambiguously) represents. 

Let us now turn to the coarse graining flow of this section. Our quantisation procedure represented the operators $\ov$ and $\widehat{e^{i \lambda n b}}$ unambiguously. We then assembled $\mathfrak{su}(1,1)$ generators from them in \eqref{eq:RegVM}, leading to the consistency requirement $\tilde v_m = 2 \lambda (j-1/2)$.
In order to apply the coarse graining results from section \ref{sec:CoarseGraining} also here, we can proceed as follows. 
\begin{enumerate} 
	\item Start with an operator $\hat O^{(j_0)}$ on some $\mathfrak{su}(1,1)$ representation space with label $j_0$ embedded in the LQG Hilbert space as above. This is a copy of the fine operator acting on a single fine cell which we want to coarse grain by joining $N$ identical cells. 
	\item Write $\hat O^{(j_0)} = a \oj^{(j_0)} + b \okp^{(j_0)} + c \okm^{(j_0)}, ~ a,b,c \in \mathbb C$ as a linear combination of $\mathfrak{su}(1,1)$ generators in representation $j_0$. An important remark at this point concerns to what operators our coarse graining is applicable. While it was shown in \cite{BodendorferCoarseGrainingAs} (and cited in section \ref{sec:CoarseGraining} from there) to apply to arbitrary powers of $\oj, \okp,$ and $\okm$, it is in fact more widely applicable, in particular to (powers of) linear combinations of generators as here. This more general coarse graining will be studied further in \cite{BHAddendum}. 
	\item The coarse grained system is obtained as defined by the table in section \ref{sec:CoarseGraining}, that is $\mathfrak{su}(1,1)$ generators in representation $j_0$ map to $\mathfrak{su}(1,1)$ generators in representation $j = N j_0$. Finally, they are again linearly combined to $\hat O^{(j)}=a \oj^{(j)} + b \okp^{(j)} + c \okm^{(j)}$ with the same coefficients $a,b,c$.
\end{enumerate}

As an example, we construct the coarse grained generator of dynamics, starting with the gravitational part of the Hamiltonian constraint as given in \eqref{eq:AnsatzHgMV} as our fine definition in representation $j_0 = \frac{1}{2}$. Note that for $j_0=\frac{1}{2}$, $\tilde v_m = 0$, so that the operator reduces to \eqref{eq:HV}. Following the above recipe, we obtain by using \eqref{eq:SU11Polym} and \eqref{eq:RegVM} 
\begin{align}
	&\hat H_g^{(j)} = \frac{1}{2 \lambda} (\okx^{(j)}-\oj^{(j)}) \label{eq:HImproved} \\
	=& -\frac{1}{2 \lambda^2} \osin ~\sqrt{\hat v^2 - 4 \lambda^2 (j-1/2)^2} ~ \osin \nonumber \\
	&+  \frac{1}{8 \lambda^2} \BR{\sqrt{(\hat v +  \lambda)^2 - 4 \lambda^2 (j-1/2)^2}+\sqrt{ (\hat v -  \lambda)^2 - 4 \lambda^2 (j-1/2)^2}- 2 \hat v}. \nonumber
\end{align}
For eigenvalues of $\hat v$ much larger than the minimal eigenvalue $2 \lambda j$, the last line approaches zero and one reobtains \eqref{eq:AnsatzHgMV}. In turn, the square root in the second line is well approximated by $\hat v$ in this limit and one obtains the more common expression \eqref{eq:HV}. The differences to \eqref{eq:HV} for $j \neq 1/2$ should thus be seen as a convenient choice of ordering that allows for a very simple coarse graining operation. We also note that $\hat H_g^{(j)}$ differs from \eqref{eq:AnsatzHgMV}. \eqref{eq:AnsatzHgMV} should only be considered as an ansatz entering the construction of the algebra. While it is a well-defined operator on the LQC Hilbert space, it cannot be written as a linear combination of the $\mathfrak{su}(1,1)$ generators \eqref{eq:RegVM} and thus is not of interest of our coarse graining purposes (except for $j=1/2$, where $\tilde v_m=0$). $\hat H_g^{(j)}$ on the other hand can be written as a linear combination of \eqref{eq:RegVM}, is well-defined on the sub-space of LQC Hilbert space with volume eigenvalues $\geq 2 \lambda j$, and constitutes a viable quantisation of the classical expression for $H_g$, see \eqref{eq:HolonomyCorrections}, which is why we define it to be the generator of time translations in the quantum theory. The dynamics generated at the coarse level agrees with that of the fine level due to \eqref{eq:UonCS}.

The $j$-dependence of \eqref{eq:HImproved} can thus be interpreted as a renormalisation group flow. 
Fundamental physics takes place at $j=1/2$. Coarse graining to a scale $j>1/2$, analogous to a block-spin transformation that joins $N=2j$ spins into one, introduces a non-trivial dependence of the operators on the scale, as e.g. in the generator of the gravitational dynamics \eqref{eq:HImproved}. 
Let us note again that other (inequivalent) constructions of the $\mathfrak{su}(1,1)$ operators may be possible where the coarse graining flow could look different. We merely present a concrete example that can be studied analytically.

\subsection{Renormalised vs non-renormalised dynamics} \label{sec:RvNR}

Based on the paradigm that fundamental physics should take place in the many low spin regime ($\alpha \rightarrow 1$ in the notation below) and that physics at large scales should be obtained via coarse graining, it is now possible to quantify the error made when neglecting renormalisation ($\alpha \rightarrow 0$). To this end, we compare the evolution the coherent states \eqref{eq:DefCoh} in these two scenarios. As stated in the introduction, we consider only the gravitational part of the Hamiltonian constraint and the gravitational degrees of freedom. There are two possible interpretations for this, both bearing their own merit. On the one hand, one may consider this as a toy model that highlights a general concept. On the other hand, the precise system under study here is obtained when considering gravity coupled to non-rotating dust \cite{SwiezewskiOnTheProperties} and deparametrising the time evolution w.r.t. the dust. Then, the gravitational part of the Hamiltonian constraint becomes a true Hamiltonian that generates time evolution w.r.t. dust time and encodes the (integrated) matter energy density.

It is well known that the holonomy corrections introduced in \eqref{eq:HolonomyCorrections} lead to bouncing solutions. An interesting observable to consider in this context is the critical density $\rho_\text{b}$ at which the bounce happens. It is defined as 
\be
	\rho_\text{b} := \text{max}_{t \in \mathbb R} \frac{\left| \vev{\hat H_g(t)} \right|}{\vev{\hat v(t)}} \text{.}
\ee 
Since time-evolution is generated by $\hat H_g$ in our model, only $\hat v$ has a time dependence. 

Let us first consider the maximally non-renormalised case, where $v(t) \gg v_\text{gap}  = 2 \lambda j$ for all $t$, i.e. the minimal volume is much larger than the volume gap. 
This is the regime where most computations in loop quantum cosmology are performed. 
It can be explicitly checked\footnote{We have checked numerically that the difference of both types of coherent states approaches zero in this limit. A quick argument why this should be the case proceeds as follows. Perelomov coherent states at large $j$ can be obtained via coarse graining many such states at smaller $j$. Computing probabilities at the coarse level is akin to drawing repeatedly from independent and identically distributed probability distributions at the fine level (with finite variance) and summing the results, see \eqref{eq:ProbPropagation}. By the central limit theorem, the coarse probability distribution approaches a Gaussian distribution, which is a special case of the coherent states of \cite{CorichiCoherentSemiclassicalStates}. The correct phase is obtained if the coherent states are peaked on the same classical phase space points.} that in this case the coherent states \eqref{eq:DefCoh} approach the standard Gaussian ones usually used in loop quantum cosmology, see e.g. \cite{CorichiCoherentSemiclassicalStates}. 
In this regime, the full quantum dynamics can usually be well approximated by effective equations where evolution is generated classically by the effective Hamiltonian \eqref{eq:HolonomyCorrections} (see \cite{AshtekarQuantumNatureOfAnalytical} for the case of scalar field matter and \cite{RovelliWhyAreThe} for a general argument). A brief calculation\footnote{Employing the effective Hamiltonian $H_g = - \frac{1}{2\lambda^2} v \sin(\lambda b)^2$ and Poisson bracket $\{b,v\}=1$, the equation of motion for $v$ reads $\dot v = \{v, H_g\}$. Using that $H_g$ is a constant of motion, we can eliminate $b$ from this equation and solve it for \eqref{eq:SolEff}. $v(t)$ assumes its minimum at $t=t_b$, where the matter energy density takes its maximum value $\frac{|H_g|}{v(t_b)} = \frac{1}{2 \lambda^2}$ due to the time independence of $H_g$. } yields 
\be
	v(t) = 2 \lambda^2 |H_g| + \frac{1}{2} |H_g| (t-t_b)^2 \label{eq:SolEff}
\ee
from which we conclude $\rho_\text{b,eff.} =\frac{|H_g|}{v(t_b)} = \frac{1}{2 \lambda^2} $ irrespective of the value of $v(t_b)$. We will confirm below that the exact quantum dynamics reduces to this result in the appropriate limit $v(t_b)\gg v_{\text{gap}}$. 

We now turn to the exact quantum dynamics. The Hamiltonian operator generating evolution is \eqref{eq:HImproved} and we consider its action Perelomov coherent states in representation $j$.
Using the techniques of \cite{LivineGroupTheoreticalQuantization, BenAchourThiemannComplexifierIn} and the present paper, it is possible to perform an analytic computation of $\rho_\text{b}$ for arbitrary $v_b$, in particular also for the limiting case $v_\text{b} \rightarrow v_\text{gap}$, where the bounce volume (in a single cell) agrees with the volume gap. 
Still, the total volume can be arbitrarily large by patching together such cells along the lines of section \ref{subsec:CC}. 
This computation is the same for all representations $j$, i.e. it automatically takes into account that we may have already performed some coarse graining. In turn, the effective equations limit leading to \eqref{eq:SolEff} always refers to the limit where the bounce volume $v(t_b)$ is much larger than the volume gap in at the currently considered level of renormalisation, i.e. $v(t_b) \gg v_\text{gap} = 2 \lambda j$.

Due to \eqref{eq:UonCS}, time evolution of the coherent states can be transferred to the spinor $z = (z^0, \bar{z}^1)$. Using \eqref{eq:HImproved} as the renormalised Hamiltonian and \eqref{eq:SU11fund}, we get 
\begin{align}
	 \left( \begin{array}{r} z^0(t)  \\   \bar{z}^1(t)  \end{array} \right) =& \exp \left(i  t \hat H_g^{(j)}\right)  \left( \begin{array}{r} z^0(0)  \\   \bar{z}^1(0)  \end{array} \right) = \exp \left(\frac{i t}{2 \lambda} (\hat k_x - \hat j_z ) \right)  \left( \begin{array}{r} z^0(0) \nonumber  \\   \bar{z}^1(0)  \end{array} \right) \\ =& \left( \begin{array}{r} z^0(0) - \frac{i t}{4\lambda} \left(z^0(0)-\bar{z}^1(0) \right) \\   \bar{z}^1(0) - \frac{i t}{4\lambda} \left(z^0(0)-\bar{z}^1(0) \right)  \end{array} \right) \label{eq:EvolutionZ}
\end{align}
It is convenient to perform the variable transformation $x = z^0 - \bar{z}^1$, $y=z^0 + \bar{z}^1$, leading to $x(t)=x(0)$ and $y(t) = y(0)- \frac{i t}{2 \lambda} x(0)$. Standard results for the expectation values of $\hat j_z$, $\hat k_x$, and $\hat k_y$ in the coherent states \eqref{eq:DefCoh}, see e.g. equations (3.3)-(3.5) in \cite{BodendorferCoarseGrainingAs}, lead to
\begin{align}	
	\vev{\hat H_g^{(j)}} &= \braopket{j,z(t)}{ \frac{1}{2 \lambda} (\okx^{(j)}-\oj^{(j)})}{j,z(t)}  =- \frac{j}{2 \lambda} \frac{(z^0-\bar{z}^1)\overline{(z^0-\bar{z}^1)}}{|z^0|^2-|z^1|^2}= - \frac{j}{2 \lambda} \frac{1}{\alpha} ,\\ & \nonumber \\
	\vev{\hat v(t)} &= \braopket{j,z(t)}{ {2 \lambda} \oj^{(j)}}{j,z(t)} = 2 \lambda j \frac{|z^0|^2+|z^1|^2}{|z^0|^2-|z^1|^2} \nonumber \\  \nonumber \\
	&= 2 \lambda^2 \left| \vev{\hat H_g^{(j)}}  \right| (1+\alpha^2) + \frac{1}{2} \left| \vev{\hat H_g^{(j)}}  \right| (t-t_b)^2,\label{eq:VQuantum} \\ \nonumber
	\\ \vev{\hat v(t_\text{b})} &=\text{min}_{t\in \mathbb R} \vev{\hat v(t)}=  2 \lambda^2 \left| \vev{\hat H_g^{(j)}}  \right| (1+\alpha^2), \\ \nonumber
	 \\ \rho_\text{b} &= \frac{\left| \vev{\hat H_g^{(j)}} \right|}{\vev{\hat v(t_\text{b})}} = \frac{1}{2 \lambda^2} \frac{1}{1+\alpha^2} \leq \frac{1}{2 \lambda^2} \text{,} \label{eq:CritDensityFromQuantum}
\end{align}
where we abbreviated $\alpha = \text{Re} \left( \frac{y(0)}{x (0)} \right)$ and $t_b=2 \lambda \text{Im}\left(\frac{y(0)}{x(0)}\right)$ minimises $\vev{\hat v(t)}$ over the evolution \eqref{eq:EvolutionZ}. We note that the expression $|z^0|^2-|z^1|^2$ appearing in all expectation values is time independent, since it is preserved, by definition of SU$(1,1)$, by SU$(1,1)$ transformations. Together with $x(t)=x(0)$, it immediately follows that $\vev{\hat H_g^{(j)}}$ is time-independent, which serves as a cross-check of the calculation.

\begin{figure}[h!]
	\centering
		\includegraphics[width=0.6\textwidth]{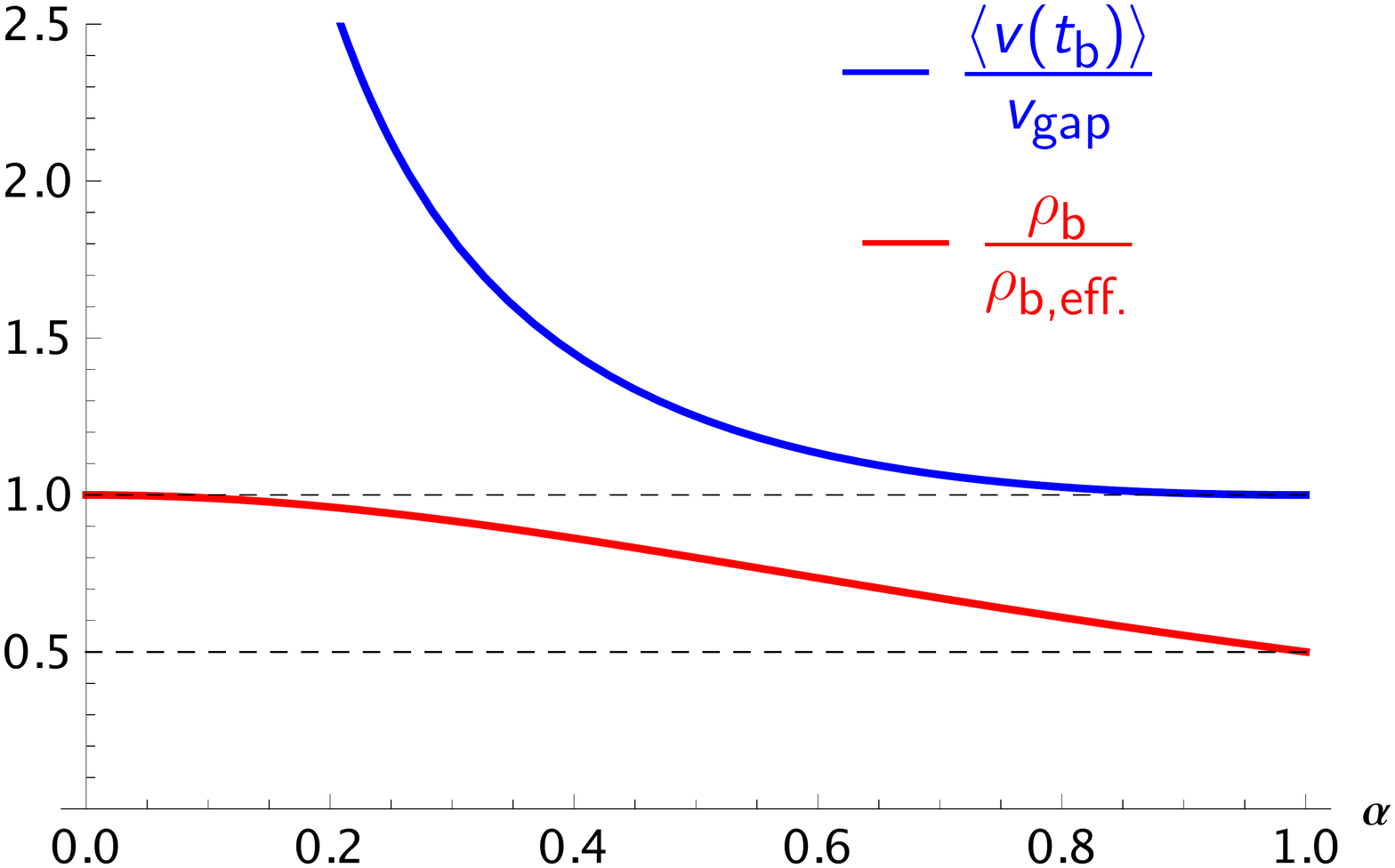}

	\caption{Dependence of the critical density $\rho_\text{b}$ on the bounce volume. The physical ``many low spins'' regime is realised for $\alpha \rightarrow 1$, where the volume at the bounce is close to the volume gap, while the ``few large spins'' regime is approached for $\alpha \rightarrow 0$, where the bounce volume is much larger than the volume gap. In the ``many low spin'' regime, the critical density approaches half the value predicted by effective equations or equivalently in the ``few large spins'' regime. }\label{fig:ren}
\end{figure}

The above computation took place in a single elementary cell. To obtain the corresponding values for the complete universe, we multiply, due to homogeneity, both $\vev{\hat H_g^{(j)}}$ and $\vev{\hat v(t_\text{b})}$ by the number of cells. This leaves $\rho_b$ unaffected. For a fixed size of the total universe, say at the bounce, the value of $\vev{\hat v(t_\text{b})}$ determines the number of cells that we resolve in our description. 
$\alpha \in (0, 1)$ thus continuously interpolates between the large (in units of the volume gap) volume per resolved cell regime ($\alpha \rightarrow 0$) where the effective equations hold, and the regime where the minimal volume per resolved cell approaches the volume gap $2 \lambda j$ ($\alpha \rightarrow 1$), which we consider as the observed physical continuum regime. We note that $\rho_\text{b}$ depends explicitly on $\alpha$ and that the limiting ``many low spin'' (many small cells) value is only half of the ``few large spin'' (few large cells) value. This provides an explicit example for the error made in computing an observable using effective equations while neglecting any renormalisation effects. The situation is visualised in figure \ref{fig:ren}. Let us remark that an overestimation of the critical density by effective equations has been noted before in \cite{CorichiCoherentSemiclassicalStates, DienerNumericalSimulationsOfACosmos} and can also be inferred from the results of \cite{AshtekarRobustnessOfKey}. Let us also note that in the limit $\alpha \rightarrow 0$, the solution \eqref{eq:SolEff} to the effective equations is reproduced by \eqref{eq:VQuantum}.

For clarity, let us comment on the limit $j\rightarrow \infty$, i.e. the limit of large quantum numbers. This limit is usually considered in the LQC literature as the large volume (of the universe) limit. While it indeed corresponds to the limit where the total volume of the universe is large, it assumes that we resolve only a single cell. The standard theory of LQC, say with the (in our language non-renormalised) Hamiltonian operator \eqref{eq:HV}, is then considered as a coarse grained description of a fundamental theory, supposedly loop quantum gravity, leading to a theory well captured by effective equations, leading e.g. to \eqref{eq:SolEff}. As we have shown in this paper, this interpretation is inconsistent with thinking of LQC as a coarse grained description of a theory involving many small quantum numbers, at least if the fundamental Hamiltonian is assumed to agree with the coarse grained one, which is common practise. While it is possible to describe the universe as a single cell, one has to use a properly renormalised Hamiltonian operator such as \eqref{eq:HImproved} in this case, which captures the effect of a fundamental description involving many cells with small quantum numbers. While our results strictly hold only for the specific model considered here, it is natural to assume that they qualitatively carry over to other theories. Let us also note that states with large quantum numbers in the absence of any implied coarse graining may have an interesting physical interpretation, see section 5.1 of \cite{BNI}, however it does not correspond to the continuum limit physics we usually observe.   

We finish with a conceptually important remark already stressed in \cite{BodendorferStateRefinementsAnd}. The computation leading to \eqref{eq:CritDensityFromQuantum} took place in a single cell and the result for the coarse observables is obtained by summing over $N$ (non-interacting) cells, which does not affect ratios of extensive quantities like $\rho_\text{b}$. While variances of operators inside a single cell may be significant as compared to their expectation values if we take a small bounce volume inside a single cell, this does not imply that the coarse observables we are interested in are not behaving classically, as they should. Rather, by an appeal to the central limit theorem in addition to the observation that the single cell variances are finite, the coarse observables are sharply peaked if the system contains sufficiently many cells $N$, and thus a large total bounce volume. This is also reflected in the fact that coherent states with large representation labels, which occur from coarse graining many cells with small representation labels, become sharply peaked.

\section{Conclusion} \label{sec:Conclusion}

In this paper, we have shown that it is possible to regularise operators corresponding to the CVH algebra on the LQC Hilbert space such that they satisfy the $\mathfrak{su}(1,1)$ Lie algebra. This result extends previous investigations \cite{LivineGroupTheoreticalQuantization, BenAchourThiemannComplexifierIn, BenAchourProtectedSL2RSymmetry} in which one started from a classical Poisson algebra with holonomy corrections. Due to extensive factor ordering problems, this result is non-trivial.  

Based on the results of \cite{BodendorferCoarseGrainingAs}, it was then possible to define and explicitly perform a coarse graining operation in a system of $N$ identical but independent copies of the same quantum cosmological system, each described by a Perelomov coherent state with labels $j_0, z$. The result can be interpreted as a non-trivial renormalisation group flow from a volume scale $j_0$ to $j=Nj_0$.

The observed difference to the coarse graining procedure in section \ref{sec:CoarseGraining}, i.e. the corrections from the naive extensive scaling, are not of much concern as they do not affect the applicability of the results of \cite{BodendorferCoarseGrainingAs}. Rather, they show that the coarse grained operators in section \ref{sec:Main} are not naively given by the quantisation of the classically coarse grained objects, but contain quantum corrections. Clearly, such effects are expected and present in most systems. One should therefore interpret section \ref{sec:CoarseGraining} as an atypical example where the coarse grained operators can be obtained naively. 
When comparing the LQC Hilbert space to that of full LQG, e.g. via an embedding along the lines of \cite{BIII, BVI}, the magnetic quantum number $m$ in the context of $\mathfrak{su}(1,1)$ is analogous to a U$(1)$ representation label, and thus to an SU$(2)$ spin $j_{\text{SU}(2)}$. In contrast, the $j$ labelling the $\mathfrak{su}(1,1)$ representation functions as a lower cutoff for $m$, and thus the smallest resolved scale set by the SU$(2)$ spins via the geometric operators. This strengthens the above interpretation of the $\mathfrak{su}(1,1)$ $j$ as a renormalisation scale. 

We have also explicitly compared the physics of the renormalised and non-renormalised models at the example of the critical density. It was found that the critical density is overestimated by a factor of two when using effective equations which correspond to the non-renormalised case ($\alpha \rightarrow 0$). 
It is important to note that this ($\alpha \rightarrow 0$) is the regime where most computations in LQC take place, i.e. one uses coarse grained quantum states that concentrate the total volume of the universe in a single vertex (in the language of the full theory embeddings \cite{BIII, BVI} of LQC), while acting on these states with the fundamental non-renormalised Hamiltonian. If one instead wants to correctly capture the physics of an underlying quantum gravity theory involving many small chunks of geometry, one needs to employ a renormalised Hamiltonian such as \eqref{eq:HImproved}. The same criticism applies to computations within full loop quantum gravity (LQG) \cite{AlesciQuantumReducedLoopGravitySemiclassicalLimit, DaporCosmologicalEffectiveHamiltonian} or related models \cite{BIII, BVI}, where one acts with the fundamental non-renormalised Hamiltonian on coarse quantum states, as well as to spin foam calculations in the same spirit, such as \cite{HanAsymptoticsOfSpinfoam}.

While our results were derived only for a specific simple model, we expect them to be valid qualitatively also in more complicated systems: it would come as no surprise that the Hamiltonians in LQG would require renormalisation when changing the resolution scale of the quantum geometry. As for the relevance of our results for applications of LQC or LQG to observations, it is important to note that our results do not show that Hamiltonians of the type \eqref{eq:HV} do not give viable phenomenology when acting on coarse quantum states. Rather, we show that if \eqref{eq:HV} is the fundamental Hamiltonian acting at the Planck scale, then the coarse dynamics is not captured by \eqref{eq:HV} acting on coarse states, but by \eqref{eq:HImproved}. Since Hamiltonians of the type \eqref{eq:HV} are usually considered as derived at the Planck scale, see \cite{ThiemannQSD1} and the application thereof in \cite{AshtekarQuantumNatureOf}, our results at least point out a conceptual flaw in many approaches relating such phenomenological models to fundamental LQG.

For future work, it is of obvious interest to check to which extend the computation performed in this paper can be generalised to more complicated systems. First, one may be interested in matter coupled to the gravitational sector. 
When quantising also the matter content, one can run into the problem that the classical value of Casimir operator is zero or negative, see e.g. \cite{BenAchourThiemannComplexifierIn}, which selects different classes of $\mathfrak{su}(1,1)$ representations, so that the analysis of \cite{BodendorferCoarseGrainingAs} would have to be successfully repeated for these cases. 
Another route is to identify suitable $\mathfrak{su}(1,1)$ sub-algebras in increasingly complicated systems, such as spherical symmetry. 

\section*{Acknowledgments}

NB was supported by an International Junior Research Group grant of the Elite Network of Bavaria and would like to thank Parampreet Singh for email exchanges on the topic of this paper, as well as Fabian Haneder for discussions about coarse graining.

\begin{appendix}

\section{Elements of $\mathfrak{su}(1,1)$ representation theory} \label{sec:AppSU11}

In this appendix, we review elements of the representation theory of SU$(1,1)$ and its Lie algebra $\mathfrak{su}(1,1)$, see \cite{SchliemannCoherentStatesOf, RamondBook} for additional details. 
The Lie algebra $\mathfrak{su}$(1,1) with generators $\okx, \oky, \oj$ features the commutation relations
\be
\BS{\okx, \oky} = -i \oj \quad \BS{\oky, \oj} = i \okx \quad \BS{\oj, \okx} = i \oky 
\ee
which can also be expressed by defining $\okpm = \okx \pm i \oky$ as
\be
 \BS{\okp , \okm} = -2 \oj \quad \BS{\oj, \okpm} = \pm \okpm \text{.}
\ee
The Casimir operator is given by
\begin{align}
\casi = \oj^2 - \okx^2 - \oky^2 = \oj^2 \mp \oj - \okpm \okmp \text{.}
\end{align}
This algebra is a non-compact form of $\mathfrak{su}$(2). As a result, all unitary irreducible representation of $\mathfrak{su}$(1,1) are infinite dimensional. These representations are labeled by the representation label $j$ determining the action of the Casimir operator and the eigenstates are labeled by the eigenvalue $m$ of $\oj$:
\begin{align}
\casi \left| j,m\right> = j(j-1)\left| j,m\right> , ~~\qquad \oj \left| j,m\right> = m \left| j,m\right> \text{.}
\end{align} 
By using the commutation relations, the action of the ladder operators $\okpm$ can be obtained as
\begin{align}
\okpm \left|j,m\right> =  \sqrt{m(m\pm1)-j(j-1)} \left|j,m \pm 1\right> \text{.} \label{eq:ActionK-}
\end{align}

There exist five possible groups of such representation. First, one can distinguish two classes, the continuous and the discrete one. We will not discuss the continuous ones here. The three discrete ones are given by  
\begin{itemize}
\item $j = \frac{1}{2},  1, \frac{3}{2}, \ldots \quad m = j, j+1, j+2, \ldots$
\item $j = \frac{1}{2},  1, \frac{3}{2}, \ldots \quad m = -j, -j-1, -j-2, \ldots$
\item $j =\frac{1}{4}, \frac{3}{4} \quad \quad ~~~~~ m \in j, j+1, j+2, \ldots$
\end{itemize} 
The last one can be obtained as the infinite dimensional Hilbert space generated by a bosonic harmonic oscillator. For our analysis, the first two cases are of interest. We restrict to positive $m$, which is case one, due to the interpretation of $2 \lambda m$ as the volume. As one can easily see, the action of $\casi$, $\oj$ and $\okpm$ always results in new states belonging to this representation. The action of $\okm$ on the lowest eigenstate $\left|j,j\right>$ vanishes.

Beside those infinite dimensional representations, also finite dimensional (non unitary) representations exit. Their dimension is, similar to the dimensions of the representations of $\mathfrak{su}$(2), given by $ 2j +1$. To find the generators, one we can simply take the $\mathfrak{su}(2)$ generators and multiply two of them by $i$, which yields for the two-dimensional defining representation 
\begin{align}
\begin{split}
\oj = \frac{1}{2}\left(\begin{array}{cc} 1 & 0 \\
	  0 & -1\end{array}\right) ,
	  \qquad \okx  = 
	  \frac{1}{2}\left(\begin{array}{cc} 0 & 1 \\
	  -1 & 0\end{array}\right),
	   \qquad \oky  =  \frac{-i}{2}\left(\begin{array}{cc} 0 & 1 \\
	  1 & 0\end{array}\right) \text{.}
	  \label{eq:SU11fund}
\end{split}
\end{align} 
acting on the spinors $(z^0, \bar{z}^1)$. The exponentiated action $U = e^{i\alpha^i \sigma_i}$ preserves the pseudo-norm $|z_0|^2 - |z_1|^2$.

\end{appendix}

%\bibliographystyle{utphysmendeley}
%\bibliography{library}

\begin{thebibliography}{10}

\bibitem{ThiemannModernCanonicalQuantum}
T.~Thiemann, {\em {Modern Canonical Quantum General Relativity}}.
\newblock Cambridge University Press, Cambridge, 2007.

\bibitem{RovelliBook2}
C.~Rovelli and F.~Vidotto, {\em {Covariant Loop Quantum Gravity: An Elementary
  Introduction to Quantum Gravity and Spinfoam Theory}}.
\newblock Cambridge University Press, 2014.

\bibitem{MarkopoulouCoarseGrainingIn}
F.~Markopoulou, ``{Coarse graining in spin foam models},'' {\em Class. Quantum
  Gravity} {\bf 20} (2003) 777--799, {\tt arXiv:gr-qc/0203036}.

\bibitem{OecklRENORMALIZATIONFORSPIN}
R.~Oeckl, ``{RENORMALIZATION FOR SPIN FOAM MODELS OF QUANTUM GRAVITY},'' in
  {\em Tenth Marcel Grossmann Meet.}, pp.~2296--2300, World Scientific
  Publishing Company2006.
\newblock {\tt arXiv:gr-qc/0401087}.

\bibitem{LivineCouplingOfSpacetime}
E.~R. Livine and D.~Oriti, ``{Coupling of spacetime atoms in 4D spin foam
  models from group field theory},'' {\em J. High Energy Phys.} {\bf 2007}
  (2007) 92, {\tt arXiv:gr-qc/0512002}.

\bibitem{DittrichCoarseGrainingMethods}
B.~Dittrich, F.~C. Eckert, and M.~Martin-Benito, ``{Coarse graining methods for
  spin net and spin foam models},'' {\em New J. Phys.} {\bf 14} (2012) 035008,
  {\tt arXiv:1109.4927 [gr-qc]}.

\bibitem{BahrHolonomySpinFoam}
B.~Bahr, B.~Dittrich, F.~Hellmann, and W.~Kaminski, ``{Holonomy spin foam
  models: Definition and coarse graining},'' {\em Phys. Rev. D} {\bf 87} (2013)
  044048, {\tt arXiv:1208.3388 [gr-qc]}.

\bibitem{BahrOnBackgroundIndependent}
B.~Bahr, ``{On background-independent renormalization of spin foam models},''
  {\tt arXiv:1407.7746 [gr-qc]}.

\bibitem{BahrHypercuboidalRenormalizationIn}
B.~Bahr and S.~Steinhaus, ``{Hypercuboidal renormalization in spin foam quantum
  gravity},'' {\em Phys. Rev. D} {\bf 95} (2017) 126006, {\tt arXiv:1701.02311
  [gr-qc]}.

\bibitem{BahrNumericalEvidenceFor}
B.~Bahr and S.~Steinhaus, ``{Numerical evidence for a phase transition in 4d
  spin foam quantum gravity},'' {\tt arXiv:1605.07649 [gr-qc]}.

\bibitem{CarrozzaFlowingInGroup}
S.~Carrozza, ``{Flowing in Group Field Theory Space: a Review},'' {\em
  Symmetry, Integr. Geom. Methods Appl.} (2016) {\tt arXiv:1603.01902 [gr-qc]}.

\bibitem{BahrRenormalizationOfSymmetry}
B.~Bahr, G.~Rabuffo, and S.~Steinhaus, ``{Renormalization of symmetry
  restricted spin foam models with curvature in the asymptotic regime},'' {\em
  Phys. Rev. D} {\bf 98} (2018) 106026, {\tt arXiv:1804.00023 [gr-qc]}.

\bibitem{DittrichCoarseGrainingFlow}
B.~Dittrich, E.~Schnetter, C.~J. Seth, and S.~Steinhaus, ``{Coarse graining
  flow of spin foam intertwiners},'' {\em Phys. Rev. D} {\bf 94} (2016) 124050,
  {\tt arXiv:1609.02429 [gr-qc]}.

\bibitem{BodendorferStateRefinementsAnd}
N.~Bodendorfer, ``{State refinements and coarse graining in a full theory
  embedding of loop quantum cosmology},'' {\em Class. Quantum Gravity} {\bf 34}
  (2017) 135016, {\tt arXiv:1607.06227 [gr-qc]}.

\bibitem{BodendorferCoarseGrainingAs}
N.~Bodendorfer and F.~Haneder, ``{Coarse graining as a representation
  change},'' {\em Phys. Lett. B} {\bf 792} (2019) 69--73, {\tt arXiv:1811.02792
  [gr-qc]}.

\bibitem{ThiemannRenormalisationReview}
T.~Thiemann, ``{Canonical Quantum Gravity, Constructive QFT and
  Renormalisation},'' {\tt arXiv:2003.13622 [gr-qc]}.

\bibitem{HanSpinfoamsNearA}
M.~Han and M.~Zhang, ``{Spinfoams near a classical curvature singularity},''
  {\em Phys. Rev. D} {\bf 94} (2016) 104075, {\tt arXiv:1606.02826 [gr-qc]}.

\bibitem{GielenCosmologyFromGroup}
S.~Gielen, D.~Oriti, and L.~Sindoni, ``{Cosmology from Group Field Theory
  Formalism for Quantum Gravity},'' {\em Phys. Rev. Lett.} {\bf 111} (2013)
  31301, {\tt arXiv:1303.3576 [gr-qc]}.

\bibitem{OritiEmergentFriedmannDynamics}
D.~Oriti, L.~Sindoni, and E.~Wilson-Ewing, ``{Emergent Friedmann dynamics with
  a quantum bounce from quantum gravity condensates},'' {\em Class. Quantum
  Gravity} {\bf 33} (2016) 224001, {\tt arXiv:gr-qc/1602.05881}.

\bibitem{BojowaldDynamicalCoherentStates}
M.~Bojowald, ``{Dynamical coherent states and physical solutions of quantum
  cosmological bounces},'' {\em Phys. Rev. D} {\bf 75} (2007) 123512, {\tt
  arXiv:gr-qc/0703144}.

\bibitem{BojraDynamicsForA}
E.~F. Borja, J.~Diaz-Polo, I.~Garay, and E.~R. Livine, ``{Dynamics for a
  2-vertex quantum gravity model},'' {\em Class. Quantum Gravity} {\bf 27}
  (2010) 235010, {\tt arXiv:1006.2451 [gr-qc]}.

\bibitem{LivineGroupTheoreticalQuantization}
E.~R. Livine and M.~Martin-Benito, ``{Group theoretical quantization of
  isotropic loop cosmology},'' {\em Phys. Rev. D} {\bf 85} (2012) 124052, {\tt
  arXiv:1204.0539 [gr-qc]}.

\bibitem{BenAchourThiemannComplexifierIn}
J.~{Ben Achour} and E.~R. Livine, ``{Thiemann complexifier in classical and
  quantum FLRW cosmology},'' {\em Phys. Rev. D} {\bf 96} (2017) 066025, {\tt
  arXiv:1705.03772 [gr-qc]}.

\bibitem{PerelomovCoherentStatesFor}
A.~M. Perelomov, ``{Coherent states for arbitrary Lie group},'' {\em Commun.
  Math. Phys.} {\bf 26} (1972) 222--236, {\tt arXiv:math-ph/0203002}.

\bibitem{PerelomovBook}
A.~M. Perelomov, {\em {Generalized Coherent States and Their Applications}}.
\newblock Springer, Berlin, 1986.

\bibitem{AntoineTwoDimensionalWavelets}
J.-P. Antoine, R.~Murenzi, P.~Vandergheynst, and S.~Ali, {\em {Two-Dimensional
  Wavelets and their Relatives}}.
\newblock Cambridge University Press, Cambridge (UK), 2004.

\bibitem{BrownDustAsStandard}
J.~Brown and K.~Kuchar, ``{Dust as a standard of space and time in canonical
  quantum gravity},'' {\em Phys. Rev. D} {\bf 51} (1995) 5600--5629, {\tt
  arXiv:gr-qc/9409001}.

\bibitem{SwiezewskiOnTheProperties}
J.~Swiezewski, ``{On the properties of the irrotational dust model},'' {\em
  Class. Quantum Gravity} {\bf 30} (2013) 237001, {\tt arXiv:1307.4687
  [gr-qc]}.

\bibitem{BojowaldAbsenceOfSingularity}
M.~Bojowald, ``{Absence of a Singularity in Loop Quantum Cosmology},'' {\em
  Phys. Rev. Lett.} {\bf 86} (2001) 5227--5230, {\tt arXiv:gr-qc/0102069}.

\bibitem{AshtekarMathematicalStructureOf}
A.~Ashtekar, M.~Bojowald, and J.~Lewandowski, ``{Mathematical structure of loop
  quantum cosmology},'' {\em Adv.Theor.Math.Phys.} {\bf 7} (2003) 233--268,
  {\tt arXiv:gr-qc/0304074}.

\bibitem{AshtekarQuantumNatureOf}
A.~Ashtekar, T.~Pawlowski, and P.~Singh, ``{Quantum nature of the big bang:
  Improved dynamics},'' {\em Phys. Rev. D} {\bf 74} (2006) 084003, {\tt
  arXiv:gr-qc/0607039}.

\bibitem{AshtekarLoopQuantumCosmology}
A.~Ashtekar and P.~Singh, ``{Loop quantum cosmology: a status report},'' {\em
  Class. Quantum Gravity} {\bf 28} (2011) 213001, {\tt arXiv:1108.0893
  [gr-qc]}.

\bibitem{SinghLoopQuantumCosmologyABrief}
P.~Singh and I.~Agullo, ``{Loop Quantum Cosmology: A brief review},'' {\tt
  arXiv:1612.01236 [gr-qc]}.

\bibitem{AlesciANewPerspective}
E.~Alesci and F.~Cianfrani, ``{A new perspective on cosmology in Loop Quantum
  Gravity},'' {\em Europhys. Lett.} {\bf 104} (2013) 10001, {\tt
  arXiv:1210.4504 [gr-qc]}.

\bibitem{BIII}
N.~Bodendorfer, ``{Quantum reduction to Bianchi I models in loop quantum
  gravity},'' {\em Phys. Rev. D} {\bf 91} (2015) 081502(R), {\tt
  arXiv:1410.5608 [gr-qc]}.

\bibitem{BVI}
N.~Bodendorfer, ``{An embedding of loop quantum cosmology in (b,v) variables
  into a full theory context},'' {\em Class. Quantum Gravity} {\bf 33} (2016)
  125014, {\tt arXiv:1512.00713 [gr-qc]}.

\bibitem{AssanioussiEmergentDeSitter}
M.~Assanioussi, A.~Dapor, K.~Liegener, and T.~Paw{\l}owski, ``{Emergent de
  Sitter Epoch of the Quantum Cosmos from Loop Quantum Cosmology},'' {\em Phys.
  Rev. Lett.} {\bf 121} (2018) 081303, {\tt arXiv:1801.00768 [gr-qc]}.

\bibitem{HellingHigherCurvatureCounter}
R.~Helling, ``{Higher curvature counter terms cause the bounce in loop
  cosmology},'' {\tt arXiv:0912.3011 [gr-qc]}.

\bibitem{IshamTopologicalAndGlobal}
C.~J. Isham, ``{Topological and global aspects of quantum theory},'' in {\em
  Relativ. Groups Topol. II} (B.~S. DeWitt and R.~Stora, eds.).
\newblock North-Holland, Amsterdam, 1984.

\bibitem{ThiemannComplexifierCoherentStates}
T.~Thiemann, ``{Complexifier coherent states for quantum general relativity},''
  {\em Class. Quantum Gravity} {\bf 23} (2006) 2063--2117, {\tt
  arXiv:gr-qc/0206037}.

\bibitem{GielenHomogeneousCosmologiesAs}
S.~Gielen, D.~Oriti, and L.~Sindoni, ``{Homogeneous cosmologies as group field
  theory condensates},'' {\em J. High Energy Phys.} {\bf 2014} (2014) 13, {\tt
  arXiv:1311.1238 [gr-qc]}.

\bibitem{BHAddendum}
N.~Bodendorfer and F.~Haneder, ``(to appear),''.

\bibitem{BenAchourProtectedSL2RSymmetry}
J.~{Ben Achour} and E.~R. Livine, ``{Protected SL(2,R) Symmetry in Quantum
  Cosmology},'' {\tt arXiv:1904.06149 [gr-qc]}.

\bibitem{WuhrerMasterarbeit}
D.~Wuhrer, {\em {SU(1,1) coherent states on the LQC Hilbert space}}.
\newblock Master thesis, Regensburg,
\newblock 2019.

\bibitem{CorichiCoherentSemiclassicalStates}
A.~Corichi and E.~Montoya, ``{Coherent semiclassical states for loop quantum
  cosmology},'' {\em Phys. Rev. D} {\bf 84} (2011) 44021, {\tt arXiv:1105.5081
  [gr-qc]}.

\bibitem{AshtekarQuantumNatureOfAnalytical}
A.~Ashtekar, T.~Pawlowski, and P.~Singh, ``{Quantum nature of the big bang: An
  analytical and numerical investigation},'' {\em Phys. Rev. D} {\bf 73} (2006)
  124038, {\tt arXiv:gr-qc/0604013}.

\bibitem{RovelliWhyAreThe}
C.~Rovelli and E.~Wilson-Ewing, ``{Why are the effective equations of loop
  quantum cosmology so accurate?},'' {\em Phys. Rev. D} {\bf 90} (2014) 23538,
  {\tt arXiv:1310.8654 [gr-qc]}.

\bibitem{DienerNumericalSimulationsOfACosmos}
P.~Diener, B.~Gupt, and P.~Singh, ``{Numerical simulations of a loop quantum
  cosmos: robustness of the quantum bounce and the validity of effective
  dynamics},'' {\em Class. Quantum Gravity} {\bf 31} (2014) 105015, {\tt
  arXiv:1402.6613 [gr-qc]}.

\bibitem{AshtekarRobustnessOfKey}
A.~Ashtekar, A.~Corichi, and P.~Singh, ``{Robustness of key features of loop
  quantum cosmology},'' {\em Phys. Rev. D} {\bf 77} (2008) 024046, {\tt
  arXiv:0710.3565 [gr-qc]}.

\bibitem{BNI}
N.~Bodendorfer and Y.~Neiman, ``{Imaginary action, spinfoam asymptotics and the
  'transplanckian' regime of loop quantum gravity},'' {\em Class. Quantum
  Gravity} {\bf 30} (2013) 195018, {\tt arXiv:1303.4752 [gr-qc]}.

\bibitem{AlesciQuantumReducedLoopGravitySemiclassicalLimit}
E.~Alesci and F.~Cianfrani, ``{Quantum reduced loop gravity: Semiclassical
  limit},'' {\em Phys. Rev. D} {\bf 90} (2014) 24006, {\tt arXiv:1402.3155
  [gr-qc]}.

\bibitem{DaporCosmologicalEffectiveHamiltonian}
A.~Dapor and K.~Liegener, ``{Cosmological effective Hamiltonian from full loop
  quantum gravity dynamics},'' {\em Phys. Lett. B} {\bf 785} (2018) 506--510,
  {\tt arXiv:1706.09833 [gr-qc]}.

\bibitem{HanAsymptoticsOfSpinfoam}
M.~Han and M.~Zhang, ``{Asymptotics of spinfoam amplitude on simplicial
  manifold: Lorentzian theory},'' {\em Class. Quantum Gravity} {\bf 30} (2013)
  165012, {\tt arXiv:1109.0499 [gr-qc]}.

\bibitem{ThiemannQSD1}
T.~Thiemann, ``{Quantum spin dynamics (QSD)},'' {\em Class. Quantum Gravity}
  {\bf 15} (1998) 839--873, {\tt arXiv:gr-qc/9606089}.

\bibitem{SchliemannCoherentStatesOf}
J.~Schliemann, ``{Coherent states of su(1,1): correlations, fluctuations, and
  the pseudoharmonic oscillator},'' {\em J. Phys. A Math. Theor.} {\bf 49}
  (2016) 135303, {\tt arXiv:1508.04549 [quant-ph]}.

\bibitem{RamondBook}
P.~Ramond, {\em {Group Theory. A Physicist's Survey.}}
\newblock Cambridge University Press, 2010.

\end{thebibliography}

\end{document}